\documentclass[%
 reprint,showpacs,
 amsmath,amssymb,
 aps,nofootinbib,superscriptaddress
]{revtex4-1}

 \usepackage{ulem}
    \usepackage{float}
\usepackage{graphicx}% Include figure files
\usepackage{dcolumn}% Align table columns on decimal point
\usepackage[colorlinks=true, linkcolor = red, citecolor = blue]{hyperref}
\usepackage{subcaption}
\usepackage{float}
\usepackage{xcolor}

\usepackage{kantlipsum}
\begin{document}
\newcommand{\lu}[1]{\textcolor{red}{#1}}
\newcommand{\quita}[1]{\textcolor{orange}{#1}}
\newcommand{\JCH}[1]{\textcolor{blue}{#1}}
\title{Constraining inflationary potentials with inflaton PBHs}% Force line breaks with \\
\author{Luis E. Padilla}
\email{lepadilla@icf.unam.mx}
\affiliation{Instituto de Ciencias Físicas, Universidad Nacional Autónoma de México,
62210, Cuernavaca, Morelos, México.}

\author{Juan Carlos Hidalgo}
  \email{hidalgo@icf.unam.mx}
\affiliation{Instituto de Ciencias Físicas, Universidad Nacional Autónoma de México,
62210, Cuernavaca, Morelos, México.}

\author{Gabriel German}
  \email{gabriel@icf.unam.mx}
  \affiliation{Instituto de Ciencias Físicas, Universidad Nacional Autónoma de México,
62210, Cuernavaca, Morelos, México.}

\date{\today}% 

\begin{abstract}
If, after primordial inflation, the universe undergoes a relatively long reheating period, it could present a phase of matter domination supported by the oscillating inflaton field. During this epoch, small perturbations from the inflaton that reenter the cosmological horizon could virialize to form \textit{inflaton} structures. If the primordial overdensities are large enough, their associated inflaton structures could collapse to form primordial black holes (PBHs) [L.~E.~Padilla, J.~C.~Hidalgo and K.~A.~Malik, Phys.~Rev.~D, vol.~106, p.~023519, Jul 2022; hereinafter P1]. For this to happen at a considerable rate, the primordial power spectrum should be enhanced at small scales, a feature typically induced in single-field inflation through an ultra-slow roll phase (produced by a nearly-inflection point in the inflationary potential). In this article we consider two specific inflationary potentials that present this nearly-inflection point and we look at the PBH formation rate through the mechanism proposed in P1. We report on constraints to these two specific models from the bounds to PBH abundances. This serves as an illustration of the usefulness of the PBH formation mechanism proposed in P1. 
\end{abstract}

\maketitle

\section{Introduction}

%It is now well established that the inflationary cosmology is the theory that best describes the early universe. During this primordial accelerated epoch it is usually expected that quantum fluctuations of the inflaton field were amplified to cosmological scales, causing in turn the amplification of gravitational and tensor perturbations. Observations of the CMB anisotropies and the large scale structure have shown to be in good agreement with an approximate scale invariant spectrum of small density fluctuations on large scales \citep{Kolb:1990vq}.  

In recent years high-precision astronomy and cosmology have developed rapidly, which has given way to scientists venturing and assess phenomena previously regarded as exotic. Since 2015 several gravitational wave signals have been detected and interpreted as the merger of binary black holes (BHs), previously unobserved objects and now detected in pairs \citep{PhysRevLett.116.061102,PhysRevLett.119.161101}.
%In addition, the direct image of the supermassive black hole at the center of M87 was observed for the first time by the Event Horizon Telescope not long ago \citep{Akiyama_2019}. %Just three years later the same collaboration obtained an image of Sagitarius A*, the supermassive black hole located at the center of our own galaxy \citep{EventHorizonTelescopeCollaboration_2022}. Its study has been so attractive lately that even the Nobel Prize in Physics in 2020 was awarded to studies that relied on these objects.
%At the smallest scales there is still not enough data to ensure the shape that the spectrum of perturbations should have, which could allow large deviations to the almost scale-invariant spectrum favored by the largest scales. 
The recent detection by LIGO and Virgo of a BH of intermediate mass \citep{abbott2020gw190521} (of order $10^2 M_\odot$), in a range of masses forbidden for BHs of stellar origin, has led to the interpretation that this object could be a 
primordial black hole (PBH) \citep{sasaki2016primordial}. 

Unlike stellar BHs, which mass range is bounded by the Chandrasekhar and Oppenheimer limits, PBHs can, in principle, be formed in a mass range of several orders of magnitude ($10^{-38}<M_{\rm PBH}/M_\odot< 10^{12}$). This is because their formation mechanism is completely different to that of stellar BHs. The standard mechanism of PBH formation assumes that these objects were formed when  primordial density fluctuations  that initially streched out of the cosmological horizon during inflation, reenter the cosmological horizon with a sufficiently dense profile that they collapse under their own gravity and form a black hole \citep{zel1967hypothesis,hawking1971gravitationally}. The amplitude of the overdensity required for collapse is dictated by the threshold --or critical-- value of the density contrast $\delta_c$ at horizon crossing. That is, any perturbation whose density contrast is greater than this threshold value $\delta_c$ at the time it reenters the horizon, should inevitably end up collapsing to form a PBH. In the particular case in which the universe is radiation dominated at horizon reentry, this threshold value is roughly $\delta_c^{(R)}\simeq 0.41-0.62$ (see for example \citep{harada2013threshold,PhysRevD.59.124013,IHawke_2002,Musco_2009}).

%However, PBHs have been widely studied since the last 50 years \citep{zel1967hypothesis} given that these objects could be subject to a great variety of physical phenomena. For example, PBHs with masses $M_{PBH}\leq 10^{15}~\rm{g}$ are expected to have evaporated at the present time due to Hawking radiation \citep{hawking1974black}, leaving behind a supersymetric particle or a Planck mass relic (see for example \citep{carr2010new}). In the opposite case, in which the mass of these PBHs were $M\geq 10^{15}~\rm{g}$, the Hawking radiation is not enough to evaporate them, and then, several cosmological and astrophysical phenomena are expected to occur for this class of objects (for a detailed understanding of all these effects, it is recommended to see the reviews \citep{carr2010new,carr2020constraints})

Radiation domination right after primordial inflation is not mandatory. A slow transition from the inflationary period to the so called "Hot Big Bang" stage is characteristic of a long reheating process. In that case, the inflaton field could experience fast oscillations around the minimum of its potential for a considerable time until its decay. This can occur at energy scales much smaller than those associated with the inflation mass. The only restriction to the period of fast oscillations (reheating) is that the inflaton must eventually transfer its energy content to the rest of the particles of the standard model of particle physics at times prior to Big Bang Nucleosynthesis (BBN).

The period of fast oscillations around the minimum of the potential of the inflaton field is usually approximated with a quadratic-like potential. In this case we expect the universe to show a matter dominated behaviour \citep{Khlopov:1985jw,Carr:2018nkm,Carrion:2021yeh,Hidalgo:2017dfp}, which in turn may provoke a primordial structure formation period (PSFP). In a series of recent papers \citep{Niemeyer:2019gab,PhysRevD.103.063525,Eggemeier:2021smj} the authors proposed an analogy between this PSFP and the structure formation period in the so-called scalar field dark matter (SFDM) model \cite{2014NatPh..10..496S,PhysRevD.95.043541,Urena-Lopez:2019kud,10.1007/978-3-319-02063-1_9}. %\lu{The idea to propose this analogy is due to the fact that both processes can be described with the same system of differential equations, the so-called Shchr\"odinger-Poisson system\footnote{In Ref. \citep{Musoke:2019ima} was shown that this PSFP in the post-inflationary universe can be well described by the Schr\"odinger-Poisson system.}, in addition to the fact that it has been shown that the nonlinear subhorizon evolution of cosmological perturbations of scalar fields does not depend on the details of the initial spectrum \citep{Musoke:2019ima}. Therefore, the different computational techniques in SFDM-only simulations as well as the different results that have been obtained from them can be used to describe the very early universe.} 
{Following this analogy and inspired in the results obtained in Ref. \cite{Padilla:2020sjy} (see also \citep{Avilez:2017jql}) for the formation of supermassive black holes in the SFDM model}, some of us have proposed, in \citep{Padilla:2021zgm}, P1, and subsequently in \citep{Hidalgo:2022yed} (hereinafter P2), a new mechanism for the formation of PBHs during the reheating epoch. This stipulates the gravitational collapse of the structures that form during the extended reheating era.  In P1 we determined a threshold value $\delta_c$ by which perturbations should gravitationally collapse to form PBHs, with $\delta_c^{(IH)} = 0.238$ the critical value for the formation of PBHs from the gravitational collapse of an \textit{inflaton halo} (IH), and $\delta_c^{(IS)} = 0.019$ in the case of PBHs collapsed from the halo core, or  \textit{inflaton star} (IS), which results from the formation of a solitonic configuration at the centre of virialized haloes, if the PSFP lasts long enough. In P2 we extended this study by considering a toy model of a generic primordial power spectrum (PPS) with a Gaussian peak in the smallest scales. In P2 we looked at the requirements and some implications for the realization of this mechanism of PBH formation. In particular, we found that, if reheating last long enough, a peak on the smallest scales would produce an excess of PBHs due to the collapse of the inflaton stars.

The main motivation of the present work is to continue the line of thought of papers P1 and P2 by looking at realizations of this new mechanism to specific inflationary models. While in P2 we worked out a generic PPS with a peak added by hand at the suitable small scales, in this article we show how such peaks can be obtained from explicit, well motivated, inflationary models and explore to which extent these models can be constrained when PBH are formed via the described mechanism. In particular, we look at specific models which present a phase of ultra slow-roll (USR) near the end of inflation. We invoke this USR phase since it generically produces peaks in the PPS in single field inflation models. Typically, such USR phase is reached when the inflaton field approaches a nearly-inflection point in its potential \citep{Tsamis:2003px,Kinney:2005vj,Iacconi_2022}. The duration of the USR phase and the location of the nearly-inflection point define in turn the location and amplitude  of the peak in the PPS. In particular, we focus on scenarios in which the USR phase affects perturbations that reenter the horizon during reheating.

The paper is organized as follows. In Sec.~\ref{secII} we present the two inflationary models studied for PBH formation. The typical CMB inflationary observables are shown in Sec.~\ref{secIII}, where we verify that both models comply with said observables. We then introduce the phase of primordial structure formation in Sec.~\ref{Sec:reheating} and we provide the conditions for primordial structures, both IHs and ISs, to collapse onto PBHs. Then, in Sec.~\ref{secV} we calculate the abundance of PBHs for each of our inflationary models, and finally in Sec.~\ref{secVI} we discuss the different constraints relevant to the resulting PBH mass spectra. In Sec.~\ref{secVII} we provide some concluding remarks.

\section{The inflationary models}\label{secII}

\begin{figure}
    \centering
    \includegraphics[width=3in]{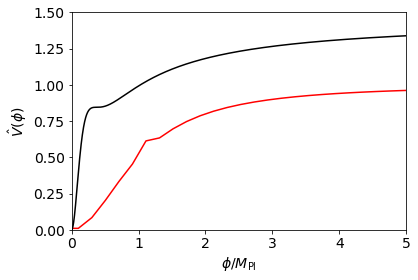}
    \caption{\footnotesize{Normalized potential $\hat V(\phi)\equiv V(\phi)/V_0^i$ ($i=I,II$) as a function of the normalized inflaton $\phi/M_{\rm Pl}$ for each of the featured models. We use the parameters given in Tables \ref{tab:1} and \ref{tab:2}. The red (black) lines corresponds to model I (II).}}
    \label{fig:potetials}
\end{figure}

In the standard single field inflationary scenario, inflation is sourced by a canonincal real scalar field $\phi$ minimally coupled to gravity and with a suitable potential $V(\phi)$. The shape of this potential is of special importance since all inflationary observables can be rewritten in terms of it and its field derivatives. In this work we  concentrate in the following explicit, complete forms, of the inflationary potential:
\begin{subequations}
\begin{itemize}
    \item Model I:
    \begin{equation}\label{eq:pot1}
    V(\phi) = V_0^{I}\frac{\phi^2}{M^2+\phi^2}\left[1+A\exp\left(-\frac{(\phi-\phi_0)^2}{2\sigma^2}\right)\right],
\end{equation}
\item Model II:
\begin{equation}\label{eq:pot2}
    V(\phi) = V_0^{II}\frac{\phi^2}{\nu^2}\frac{6-4a\frac{\phi}{\nu}+3\frac{\phi^2}{\nu^2}}{\left(1+d\frac{\phi^2}{\nu^2}\right)^2},
\end{equation}
\end{itemize}
\end{subequations}
where the parameters of each model are determined as follows: $V_0^{I}$ and  $V_0^{II}$ are set by the amplitude of the PPS at the pivot scale. $\phi_0$ is the field value at the feature in the potential of Model I, while for Model II such feature is determined by a combination of $a$, and $d$ (see Section II of Ref.~\cite{Garcia-Bellido:2017mdw}). Additionally, $\sigma$ defines the width of the ultra slow-roll section in Model I. Finally, $A$ sets the amplitude of the peak in the power spectrum feature, and $M$ is a free parameter of the theory which we set to unity. All of the above parameters are thus constant values that can be adjusted through cosmological observations. However, small variations in the parameter values change the amplitude of the power spectrum feature sensibly. In this paper we work solely with the parameter values shown in Tabs. \ref{tab:1} and \ref{tab:2}. These potentials have the particularity of having a plateau for large values of the field, required from observations of the CMB and large scale structure, while for small values, close to the minimum of the potential, the inflaton would experience a new plateau (see Fig. \ref{fig:potetials}). In particular, the inflaton potential in Model I was inspired in string theory based KKLT model \citep{Kachru:2003aw,Kachru:2003sx}. In \citep{Mishra2020PrimordialBH} the authors proposed the form in Eq. \eqref{eq:pot1} to study the effects in PBH formation from a tiny bump/dip in the inflaton potential. On the other hand, the particular potential given in Model II, Eq. \eqref{eq:pot2}, was proposed in \citep{Garcia-Bellido:2017mdw} precisely to show that a second plateau in the inflaton field has important consequences in the formation of PBHs. We must emphasize     that unlike the previously mentioned articles, our work contemplates a new mechanism for the formation of PBHs, which was not explored by the authors of said works. However, our article is similar in spirit to theirs. 
\begin{table}[h!]
    \centering
    \begin{tabular}{||c|c|c|c|c|c||}
    \hline
    Model&$V_0^I$&A&M& $\phi_0$ & $\sigma^2$ \\
    \hline
    I&$2.0078\times 10^{-10}$&0.11031&1&1.105&$5\times 10^{-3}$ \\
    \hline
    \end{tabular}
    \caption{\footnotesize{Parameters used for Model I. }}
    \label{tab:1}
    \centering
    \begin{tabular}{||c|c|c|c|c||}
    \hline
    Model&$V_0^{II}$&$\nu$&$a$ &$d$ \\
    \hline
    II&$2.59832\times 10^{-10}$&$0.3148$&1&1.43507 \\
    \hline
    \end{tabular}
    \caption{\footnotesize{Parameters used for Model II.}}
    \label{tab:2}
\end{table}

\begin{figure}
    \centering
    \includegraphics[width=3.5in]{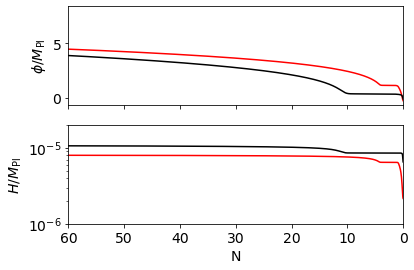}
    \caption{\footnotesize{Background evolution of the inflaton field (upper) and the Hubble parameter (lower) for each of our models and the same parameters as in Fig.~\ref{fig:potetials}. We show only the evolution of the last 60 $e$-folds before reaching the end of the inflationary period. The red (black) lines corresponds to model I (II).}}
    \label{fig:background}
\end{figure}

\begin{figure}
    \centering
    \includegraphics[width=3.5in]{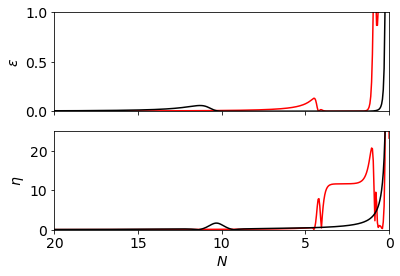}
    \caption{\footnotesize{Evolution of the slow-roll parameter for each of our models and the same parameters as in Fig.~\ref{fig:potetials}. In the figure we have only plotted the last 20 $e$-folds of inflation. The red (black) line corresponds to model I (II). }}
    \label{fig:en}
\end{figure}

Before proceeding with this work, let us discuss the implications of the two plateau regions as illustrated in Fig.~\ref{fig:potetials}. At the background level, the dynamics of the universe is described by the Friedmann and Klein-Gordon system of equations
\begin{subequations}
\begin{equation}
    H^2 = \frac{1}{3M_{\rm Pl}}\left(\frac{1}{2}\dot\phi^2+V(\phi)\right),
\end{equation}
\begin{equation}
    \ddot\phi +3H\dot\phi+V'(\phi) = 0,
\end{equation}
\end{subequations}
where an over-dot denotes a derivative with respect to the cosmic time, $M_{\rm Pl}$ is the Planck mass, $H$ is the Hubble parameter, and $V'(\phi)\equiv d\,V/d\phi$. The numerical solutions of the above equations for each of our models are plotted in Fig. \ref{fig:background} as a function of the $e$-folds number $dN \equiv d\ln(a)$, with $a$ the scale factor. We can understand the figure in the following way: during the first plateau phase, it is fulfilled the slow-roll condition\footnote{\label{footnote1}The standard slow-roll inflationary epoch takes place as long as the two slow-roll conditions ($\epsilon\equiv (1/2)\cdot[V'(\phi)/V(\phi)]^2\ll 1$ and $\eta\equiv |V^{''}(\phi)/V(\phi)|\ll 1$) are fulfilled.} ($\epsilon \ll 1$ and $\eta\ll 1$) and then the inflaton slowly rolls down the slope of its potential. During this period, the typical slow-roll inflation takes place. 
In Fig. \ref{fig:en} we can see that this condition is fulfilled approximately up to about 5-10 $e$-folds before inflation ends. At some point between the two plateaus, the kinetic term of the inflaton starts to grow and becomes maximum. Then, when the inflaton reaches the second plateau, its acceleration quickly increases and becomes comparable to the gradient of the potential. In this case the dynamics of the inflaton field can be approximated as
\begin{equation}
    \ddot\phi +3H\dot\phi = -\frac{dV(\phi)}{d\phi}\simeq 0.
\end{equation}
This phase is the well-known USR phase.\footnote{In an exact inflection point we have $\eta = 3$, whereas in a nearly-inflection point it is fulfilled the condition $\eta\gtrsim 3$ \citep{DIMOPOULOS2017262,Iacconi_2022}.} Finally, the inflationary process ends when the inflaton escapes from the second plateau and $\epsilon\simeq 1$.

During the second plateau phase it is usually expected that the perturbations $\delta\phi$ produced by the inflaton gets amplified, which in turns should induce a similar effect for scalar perturbations in the metric \citep{motohashi2017primordial}. In general, in order to visualize correctly this effect, the Mukhanov-Sasaki equations must be solved \citep{Sasaki:1986hm,Mukhanov:1988jd}:
\begin{equation}
    u_{k}^{''}+\left(k^2-\frac{z^{''}}{z}\right)u_{k} = 0,
\end{equation}
where here a prime ( $'$ ) denotes a derivative with respect to the conformal time $d\tau = dt/a$, $z\equiv a\,\dot\phi_b/H$; $u_k$ is known as the Mukhanov-Sasaki variable, and the subindex $b$ is used to refer to background quantities. 

The comoving curvature perturbation $\mathcal{R}_{k}$ is defined in term of $u_{k}$ as
\begin{equation}
    \mathcal{R}_{k}\equiv \frac{u_{k}}{z}.
\end{equation}
This quantity allows us to define the PPS of curvature perturbations as
\begin{equation}
    \mathcal{P}_{\mathcal{R}}(k)\equiv \left.\frac{k^3}{2\pi^2}|\mathcal{R}_{k}|^2\right|_{k\ll aH}.
\end{equation}
When the scale $k$ is deep inside the Hubble horizon ($k\gg aH\ \Rightarrow \ k\gg z^{''}/z$), the evolution of $u_k$ can be expressed as $u_k = \frac{1}{\sqrt{2\pi}}e^{-ik\tau}$, which represents the Bunch-Davies vacuum. On the other hand, in a quasi-de Sitter space and for scales much larger than the horizon ($k\ll aH\ \Rightarrow \ k\ll z^{''}/z$), we can solve exactly the Mukhanov-Sasaki equation, allowing us to write
\begin{equation}\label{pps}
    \mathcal{P}_\mathcal{R}(k) = \frac{H^2}{8\pi^2\epsilon}.
\end{equation}
This approximate solution to the PPS of curvature perturbations is valid beyond the slow-roll approximation, in the sense that it is derived without neglecting the acceleration and kinetic parts of the inflaton field. However it tends to underestimate the amplitude of the peak in the PPS in scenarios like the USR phase (see for example \citep{dalianis2019primordial,Bhatt:2022mmn}). As argued in \citep{dalianis2019primordial} (see the discussion in their Fig.~7) the underestimation of the amplitude of the PPS in the above formula can be attributed to the fact that the more strongly the slow-roll condition is broken in the parameter $\epsilon$, the greater is the underestimation of the PPS with the approximation formula \eqref{pps}. In this work we will only use scenarios where the slow-roll approximation is invalid only the end of inflation (see Fig.~\ref{fig:en}), so in what follows we will work with the approximate formula \eqref{pps}.

\begin{figure}
    \centering
    \includegraphics[width=3in]{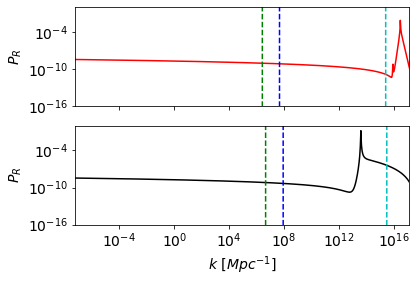}
    \caption{\footnotesize{Primordial power spectrum for each of our inflationary potentials and the same parameters as in Fig.~\ref{fig:potetials}. The red (black) lines correspond to the solution of model I (II). The green, blue, and cian dashed lines indicate the minimum k-modes that are relevant for the reheating epoch, the formation of inflaton halos, or the formation of inflaton stars, respectively (a more in depth discusion is presented below in Sec. \ref{Sec:reheating})}}
    \label{fig:spect}
\end{figure}

In Fig.~\ref{fig:spect} we plotted the PPS of curvature perturbations generated for each of our models and for the particular values of Tables \ref{tab:1} and \ref{tab:2}. Notice that, as expected and discussed previously, both PPS posses a peak at small scales as a consequence of the second plateau. However, the position of the peak in each of the models is different, this will have very important consequences regarding the formation of PBHs.  

\section{The inflationary observables}\label{secIII}

For our models to be considered a good alternative to inflationary models it is necessary that they comply with the different constraints that exist for the PPS. In particular, the observational bounds, provided by the Planck 2018 results \citep{Planck:2018jri} should be met at the scales tested by Planck. In the standard approximation what is typically constrained is the tensor-to-scalar ratio $r$ and the PPS of curvature perturbations, which can be parameterized as
\begin{equation}\label{pps1}
    \mathcal{P}_\mathcal{R}(k) = \mathcal{A}_s\left(\frac{k}{k_*}\right)^{n_s(k)-1},
\end{equation}
where $\mathcal{A}_s$ is the amplitude of the perturbations, typically quoted at the pivot scale $k_* = 0.05~\rm{Mpc^{-1}}$, and $n_s$ is known as the spectral index which, in general, depends on the scale $k$. At the lowest order in a series expansion around $k_*$ we have $n_s = {n_s|_{k_*}+...}.$ 
The constraints by Planck 2018 for these parameters are shown in Table \ref{tab:nr}. That table shows also that the parameters in Tables \ref{tab:1} and \ref{tab:2} meet the above constraints at $1\sigma$ and $2\sigma$ respectively.
\begin{table}[]
    \centering
    \begin{tabular}{|c|c|c|c|}
      \hline
         & $n_s|_{k_*}$ &$r$& $\ln(10^{10}A_s)$\\
         \hline
         Planck &$0.9649\pm 0.0042$  &$<0.036$ &$3.044\pm 0.014$  \\
         \hline
        Model I & 0.9631& 0.0055 &3.044\\
         \hline
        Model II &0.9594 & 0.0098&3.044\\
         \hline
    \end{tabular}
    \caption{\footnotesize{Numerical value of the spectral index and the tensor-to-scalar ratio at the pivot scale $k_*$ for each of our models.}}
    \label{tab:nr}
\end{table}
%\begin{figure}
%    \centering
%    \includegraphics[width = 3in]{ns_r.png}
%    \caption{\footnotesize{Spectral index (upper) and tensor-to-scalar ratio (lower) as a function of $k$ for each of our models and the parameters shown in Tabs. \ref{tab:1} and \ref{tab:2}. The blue dashed lines in the plot of the spectral index corresponds to the minimum and maximum allowed values at $1\sigma$ coming from the CMB, whereas in the plot of the tensor-to-scalar ratio is plotted the upper bound. The solid (dashed) line corresponds to model I (II).}}
%    \label{fig:ns_r}
%\end{figure}

Before discussing PBH formation, we complement this section by saying that in each of our models we have calculated numerically the time at which the pivot scale $k_*$, the scale $k_o = 10^{-4}~\rm{Mpc^{-1}}$, and the scale associated with the peak of the PPS leave the horizon, $k_{p}$. In particular, the scale $k_o$ corresponds to the largest scale currently observable and is customarily set to exit the cosmological horizon about $\sim 50-60$ $e$-folds before to end inflation, in order to solve the horizon and flatness problems. For model I (II) we found that $k_*$, $k_o$ and $k_{p}$ left the horizon $44.066$ ($43.816$), $50.304$ ($50.005$), and $1.944$ ($9.282$) $e$-folds before the end of inflation, respectively.

\section{The reheating era and formation of PBHs}\label{Sec:reheating}

The oscillating regime of the inflaton field around the minimum of its potential is typically modeled by the serie expansion
\begin{equation}
    V(\phi) \simeq \frac{1}{2}\mu^2\phi^2+...,
\end{equation}
where $\mu^2\equiv d^2V(\phi)/d\phi^2|_{\phi_{\rm min}}$ is an effective mass term of the inflaton field. The numerical values of the models we are testing in this work are shown in Table \ref{tab:mu2lambda}. In this regime, the density of the background universe is expected to go through a mattter dominated era:
\begin{equation}\label{background}
    \rho(a) \simeq \rho_{\rm end}\left(\frac{a_{\rm end}}{a}\right)^3,
\end{equation}
where subfix $ _{\rm end}$ is used for quantities evaluated at the end of inflation. This period of the universe would came to an end once the inflaton decayed and transferred its energy to other Standard Model particles. In general, the only limit to how long this period of reheating could last is that it must occurred prior to BBN. If we assume that BBN occurred at the energy density scale $\rho_{\rm BBN}\simeq (10~\rm{MeV})^4$, we can compute the maximum number of $e$-folds ($N_{\rm reh}^{\rm(max)} = (1/3)\cdot\ln(\rho_{\rm BBN}/\rho_{\rm end})$) that reheating might last for each of our models. This number is also given in Table \ref{tab:mu2lambda}.

As we discussed in P1 and P2, during this epoch some of the perturbations that left the horizon close to the end of inflation can reenter the horizon and subsequently begin a PSFP. In particular, the number of $e$-folds after inflation necessary for a $k$ scale to reenter the cosmological horizon is given by 
\begin{equation}
    N_{\rm HC}(k) = 2\ln\left(\frac{k_{\rm end}}{k}\right).
\end{equation}
After this moment, the perturbation should start to grow and become non-linear, which happens 
\begin{equation}
    N_{\rm NL}(k) = N_{\rm HC}(k)+\ln[1.39\delta_{\rm HC}^{-1}(k)]
\end{equation}
$e$-folds after the end of inflation, where $\delta_{\rm HC}(k)$ is the value of the contrast density associated to the scale $k$ and evaluated at the horizon crossing time. Then, IHs should form after
\begin{equation}
    N_{\rm IH}(k) = N_{\rm NL}(k)+\frac{2}{3}\ln\left(1+\frac{H^{-1}}{t_{\rm NL}(k)}\right)
\end{equation}
$e$-folds of expansion and with a mass of
\begin{equation}
    \left(\frac{M_{\rm IH}(k)}{7.1\times 10^{-2}~\rm{g}}\right) = \left(\frac{1.8\times 10^{15}~\rm{GeV}}{H_{\rm end}}\right)\left(\frac{k_{\rm end}}{k}\right)^3.
\end{equation}
If reheating last long enough, in the center of IHs we could expect the formation of an IS through the Bose-Einstein condensation and with a mass fulfilling the relation
\begin{equation}
    \left(\frac{M_{\rm IS}(k)}{2.4\times 10^{-5}~\rm{g}}\right) = \frac{\rho_{11}^{1/6}(a_{\rm{NL}})}{\mu_5}\left(\frac{M_{\rm IH}(k)}{7.1\times 10^{-2}~\rm{g}}\right)^{1/3}.
\end{equation}
In the above expression $\mu_5 \equiv \mu/(10^{-5}~M_{\rm Pl})$, $\rho_{11}(a_{\rm NL}) \equiv 200\rho(a_{\rm NL})/(10^{11}~\rm{GeV})^4$, and $\rho(a_{\rm NL})$ is the value of the background density evaluated at the time the perturbation $k$ becomes nonlinear. The number of $e$-folds is needed after the end of inflation for an IS to form is given by
\begin{equation}
    N_{\rm IS}(k) = N_{\rm NL}(k)+\frac{2}{3}\left(1+\frac{\Delta t_{\rm cond}(k)}{t_{\rm NL}(k)}\right),
\end{equation}
where $\Delta t_{\rm cond}(k)$ is the condensation time
\begin{equation}
    \frac{\Delta t_{\rm{cond}}(k)}{t_{\rm NL}(k)} = 8.168\times 10^{-18}\left(\frac{\mu^2}{10^{-10}M_{\rm Pl}^2}M_{\rm IH}(k)R_{\rm IH}(k)\right)^{3/2}.
\end{equation}

In P1 we show that if at the horizon crossing time the contrast density associated to the perturbations $\delta_{\rm HC}(k)$ were larger than the threshold values 
\begin{equation}
    \delta_c^{\rm (IH)} = 0.238, \ \ \ \ \ \delta_c^{\rm (IS)} = 0.019,\nonumber 
\end{equation}
we should expect that IHs or ISs collapsed to form PBHs, respectively. This means that if reheating lasts long enough, $N_{\rm reh}\geq N_{\rm IH}, N_{\rm IS}$, we could get the formation of PBHs due to the gravitational collapse of any of the 2 types of structures that could form in this period.

In Fig.~\ref{fig:spect} we sketch with a green, blue and cian vertical dashed lines the larger scale (smallest $k$) that is important for the reheating era, the formation of IHs, and the formation of ISs, respectively, and for the particular case in which $ N_{\rm reh} = 50$. We also show in Table \ref{tab:mu2lambda} the number of $e$-folds $N_{\rm HC}^{\rm peak}$, $N_{\rm IH}^{\rm peak}$ and $N_{\rm IS}^{\rm peak}$ necessary for the scale $k_{\rm peak}$ associated to the peak of the PPS to reenter the cosmological horizon, to form IHs, and to form ISs, respectively. From the table we can see that if reheating lasted at least $\sim 21$ $e$-folds we could expect that model I could form PBHs through the gravitational collapse of both class of structures, IHs and ISs. On the contrary, for model II we can see that even after 50 $e$-folds of reheating (which is very close to the the maximum value $N_{\rm reh}^{\rm (max)}$ given in Tab. \ref{tab:mu2lambda}) we would not expect the scales associated with the maximum of the PPS to reach to form ISs, so in this model we should expect that PBHs could form only by the gravitational collapse of IHs. 
\begin{table}[h!]
    \centering
\resizebox{\columnwidth}{!}{
    \begin{tabular}{||c|c|c|c|c|c|c||}
    \hline
       &$\mu^2~[10^{-10}~M_{\rm Pl}^2]$ & $N_{\rm reh}^{\rm (max)}$&$N_{\rm HC}^{\rm peak}$& $N_{\rm IH}^{\rm peak}$& $N_{\rm IS}^{\rm peak}$\\
        \hline
        Model I& 4.0157 & 54.9839& 3.699 &5.386& 20.189\\
        \hline
        Model II& 31.1798 &55.3138& 18.046&18.047&No formation\\
        \hline
    \end{tabular}
    }
    \caption{\footnotesize{We show the effective mass $\mu$, the maximum number of $e$-folds $N_{\rm reh}^{(max)}$ that reheating can last, and the number of $e$-folds $N_{\rm HC}^{\rm peak}$, $N_{\rm IH}^{\rm peak}$, and $N_{\rm IS}^{\rm peak}$ necessary for the $k_{\rm peak}$ associated to the peak of the PPS to reenter the cosmological horizon,  to form IHs, and to form ISs, respectively, for each of our models.}}
    \label{tab:mu2lambda}
\end{table}

\section{Abundance of Primordial black holes}\label{secV}
\subsection{Count of initally collapsed objects}
In the Press-Schechter formalism \citep{Press:1973iz}  the fraction of collapsed object with masses $>M$ is equivalent to the probability that an smoothed density field exceeds the threshold value $\delta_c$:
\begin{equation}\label{Pdelta}
    P[\delta>\delta_c] = \int_{\delta_c}^{\infty}P(\tilde\delta)d\tilde\delta.
\end{equation}
We can assume the probability density function associated to $\delta$, $P(\delta)$, follows a Gaussian profile
\begin{equation}
    P(\delta) = \frac{1}{\sqrt{2\pi}\sigma(R)}\exp\left(-\frac{\delta^2}{2\sigma(R)^2}\right),
\end{equation}
where $\sigma(R)$ is the variance of $\delta$ {evaluated at the horizon crossing time}, 
\begin{equation}
    \sigma^2(R
    ) = \int_0^\infty W^2(\tilde kR)\mathcal{P}_\delta(\tilde k,t_{\rm HC})d\ln \tilde k,
\end{equation}
$W(kR) = \exp(-k^2R^2/2)$ is the Fourier transform of the window function used to smooth the density contrast over a scale $R = 1/k$, and $\mathcal{P}_\delta$ is the power spectrum of density perturbations. This implies that we can rewrite Eq.~\eqref{Pdelta} as
\begin{equation}\label{Pdelta2}
    P[\delta>\delta_c] = \frac{1}{2}\text{erfc}\left(\frac{\delta_c}{\sqrt{2}\sigma(R)}\right).
\end{equation}

\begin{figure}
    \centering
    \includegraphics[width=3.5in]{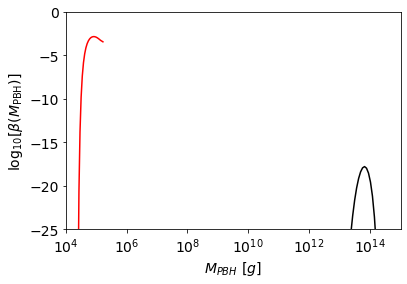}
    \caption{\footnotesize{Abundance of PBHs as a function of their mass. The red line corresponds to the abundance of PBHs for Model I, formed through the gravitational collapse of ISs. The black line shows the abundance of PBHs in Model II, formed by the collapse of IHs. }}
    \label{fig:abundance}
\end{figure}

We can finally compute the abundance  of PBHs of a given mass $M$ at the time of formation, $\beta(M)$, using the following relation:
\begin{equation}\label{betaM}
    \beta(M) = -2M\frac{\partial R}{\partial M}\frac{\partial P[\delta>\delta_c]}{\partial R},
\end{equation}
where the factor 2 is included to fit estimations from  peaks theory. We show in Fig.~\ref{fig:abundance} the abundance calculated in each of our inflationary models. As we can see in the figure, for Model I we obtain a population of PBHs that were formed due to the gravitational collapse of ISs wheras in the case of Model II we obtained a population of PBHs formed due to the gravitational collapse of IHs.

\subsection{Evolution of populations of PBHs}

Once a PBH forms it starts loosing mass via Hawking radiation. The time necessary for a PBH to evaporate completely is given by
\begin{equation}
    \Delta t_{\rm eva} \equiv t_{\rm eva}-t_f=t_{\rm Pl}\left(\frac{M_{\rm PBH}(t_{f})}{M_{\rm Pl}}\right)^{3},
\end{equation}
where $t_{\rm Pl}$ is the Planck time, $t_f$ is the formation time, and $M_{\rm PBH}(t_f)$ is the initial mass of the PBH. At the present time we expect only PBHs of mass greater than about $10^{15}~\rm{g}$ to survive. In our case, then, the PBHs generated for both models should have already evaporated. %Despite this, there are certain astrophysical and cosmological observations that would allow us to constrain our inflationary models and in this section we shall discuss such constraints.
%
%For the PBHs generated in Model I the strongest constraint comes from the Planck mass relics. 
However, in the context of quantum-gravity (see for example \citep{COLEMAN1992175}) there is the idea that BH evaporation stops when the mass of the BH reaches the Planck mass, leaving behind a relic which may contribute to the dark matter of the universe.  Following the same description as in P2, we take $\bar\beta$ as the mass fraction in absence of Hawking radiation. In this case, the evolved mass fraction of Planck mass relics at time $t$ is given by  \citep{Martin:2019nuw}:
\begin{equation}\label{omx}
    \Omega_{\rm PMR}(t) = \int_{\tilde m_{\rm min}}^{\tilde m_{\rm max}}\bar\beta(M_{\rm PBH},t)\frac{m_{\rm Pl}}{M_{\rm PBH}}d\ln M_{\rm PBH},
\end{equation}
where $\tilde m_{\rm min}$ ($\tilde m_{\rm max}$) is the minimum (maximum) mass with which PBHs originally formed and that evaporated by time $t$. As we explained in P2, $\bar \beta(M_{\rm PBH},t)$ remains constant during the reheating period and begins to evolve only when the radiation dominated era begins. By expressing $\bar\beta(M_{\rm PBH},t) = b(t)\beta(M_{\rm PBH})$, we have that $b(t)$ must fulfills the following differential equation during the radiation dominated epoch
\begin{equation}\label{eq_b}
    \frac{db}{d\ln \rho_{\rm tot}}+\frac{\Omega_{\rm PMR}-1}{\Omega_{\rm PMR}-4}b = 0,
\end{equation}
where $\rho_{\rm tot} =\rho_{\rm rad}+\rho_{\rm PMR}$ and $\rho_{\rm PMR}$ is the energy density of Planck mass relics. The system \eqref{omx} and \eqref{eq_b} is solved with appropriate initial condition such as $b(t_{\rm reh}) = 1$.
\section{Constraints to inflationary models from PBH overproduction}\label{secVI}
 We now proceed to explore the parameter values of the models of inflation exemplified above, in order to determine when the  PBH populations grow beyond the observational bounds to their abundance. Let us first describe the  observational constraints relevant to the masses of the produced PBHs. 

\begin{figure}[h!]
    \centering   \includegraphics[width=3.5in]{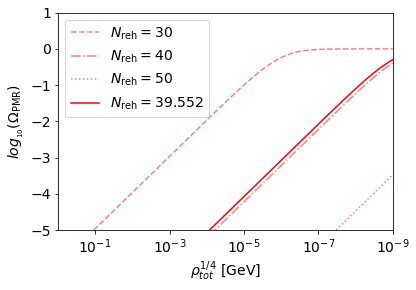}
    \caption{\footnotesize{We plotted, for Model I, the evolution of the mass fraction of Planck mass relics as a function of the energy density of the Universe after inflation. The solid red line shows the evolution that $\log_{10}(\Omega_{\rm PBH})$ must follow such that today Planck mass relics constitute all the dark matter in the universe. }}
    \label{fig:abundance_evolve}
\end{figure}

%%%PEGADO
The strongest constraint of Model I comes from requiring that Planck mass relics contribute to the totality of dark matter in the universe. In Fig.~\ref{fig:abundance_evolve} we plotted the evolution of the abundance of the Planck mass relics for Model I and for different energy scales in which reheating could have taken place. In particular, the red solid line corresponds to the number of $e$-folds that reheating would last for if the Planck remnants constitute all of the dark matter in the universe. In that case $N_{\rm reh} = 39.552$. This value is thus a lower bound on the number of $e$-folds that reheating should last for our model I, otherwise Planck mass relics would be produced in excess of the dark matter.     

On the other hand, for the spectrum of masses of PBHs in Model II the strongest constraints come from CMB and extragalactic $\gamma$-ray background (EGB) \cite{carr2020constraints}. In the standard evolution, in which the universe is radiation dominated after inflation, the CMB anisotropies constrain the abundance of PBHs in the mass range $2.5\times 10^{13}~\rm{g}\lesssim M_{\rm PBH}\lesssim 2.4\times 10^{14}~\rm{g}$ as
\begin{equation}\label{beta_1}
    \beta^{'\rm (rad)}(M_{\rm PBH})<3\times 10^{-30}\left(\frac{M_{\rm PBH}}{10^{13}~\rm{g}}\right)^{3.1},
\end{equation}
whereas the EGB constrain the mass range $M_{\rm PBH}<5\times 10^{14}~\rm{g}$ to fulfill
\begin{equation}\label{beta_2}
    \beta^{'\rm (rad)}(M_{\rm PBH})\lesssim 5\times 10^{-28}\left(\frac{M_{\rm PBH}}{5\times 10^{14}~\rm{g}}\right)^{-3.3}.
\end{equation}
In the above expression
\begin{eqnarray}
\beta^{'\rm (rad)}(M_{\rm PBH})\equiv &&\left(\frac{1}{3}\right)^{3/4}\left(\frac{g_{*i}}{106.75}\right)^{-1/4}\times\nonumber\\ 
&&\left(\frac{h}{0.67}\right)^{-2}\beta^{\rm (rad)}(M_{\rm PBH}),
\end{eqnarray}
$g_{*i}$ is the number of relativistic degrees of freedom at the formation time, $h$ is the reduced Hubble parameter, normalized to $h = 0.67$ \citep{Planck:2018jri}, and we have used the superscript $ ^{\rm (rad)}$ to refer to constraints derived in the instant reheating scenario (instant radiation domination after inflation).
It is clear that in order to constrain our Model II we need to adapt the above limits to accommodate an early matter dominated era after inflation. Let us describe the simplest way to achieve this.  

Consider for simplicity a monochromatic spectrum. In the standard -- instant reheating -- scenario, we can assume adiabatic cosmic expansion, which allow us to compute the fraction $\beta^{\rm (rad)}(M_{\rm PBH})$ of the universe collapsing into PBHs of mass $M_{\rm PBH}$ as \citep{Carr:2020gox}  
\begin{equation}
    {\displaystyle \beta^{'\rm (rad)}(M_{\rm PBH}) \simeq 7.06\times 10^{-18}\Omega^{0\rm (rad)}_{\rm PMR}\left(\frac{M_{\rm PBH}}{10^{15}~\rm{g}}\right)^{1/2}\left(\frac{M_{\rm PBH}}{M_{\rm Pl}}\right).}
\end{equation}
where $\Omega_{\rm PMR}^{0\rm (rad)}$ is the current density parameter of Planck mass relics that were originated by the evaporation of the PBHs with mass $M_{\rm PBH}$. This means that the constraints shown in Eqs. \eqref{beta_1} and \eqref{beta_2} would place a restriction on the density parameter of Planck mass relics $\Omega_{\rm PMR}^{0\rm (rad)}$ at the present time\footnote{{Even if no Planck mass relics were formed after the evaporation of the PBHs, the reader should take such abundance only as a mathematical tool that will help us to extend the aforementioned constraints to the scenario of a prolonged reheating.}}. With this idea in mind, we can then evolve Eq. \eqref{eq_b} with $\Omega(t) = \beta(M_{\rm PBH})b(t)M_{\rm Pl}/M_{\rm PBH}$ and the initial condition $b(t_{\rm reh}) = 1$, and find the initial condition in $ \beta(M_{\rm PBH})$ which would fit the constraint given by $\Omega_{\rm PMR}^{0\rm (rad)}$. Of course, in this way of finding the bounds for the abundances of PBHs in our extended reheating scenario, we are assuming that effects due to CMB and EGB occur long after BBN, so we should expect abundances in both scenarios to coincide at all times from BBN to the present time while their evolution may differ at times prior to BBN.

\begin{figure}
    \centering
    \includegraphics[width=3.5in]{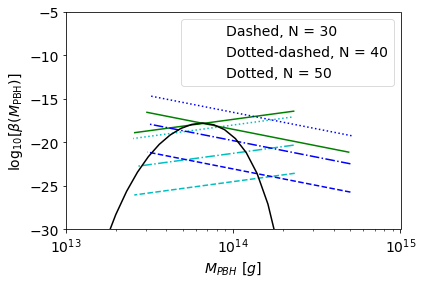}

\caption{\footnotesize{Abundance of PBHs as a function of their mass. We included the constraints from CMB anisotropies (cian) and EGB (blue). As in Fig.~\ref{fig:abundance_evolve} the dashed, dotted-dashed, and dotted lines correspond to $N_{\rm reh} = 30,\ 40,\ 50$, respectively. The green line corresponds to $N_{\rm reh} = 52$.} }
    \label{fig:my_label}
\end{figure}

In Fig.~\ref{fig:my_label} we plotted the abundance of PBHs $\beta(M_{\rm PBH})$ as a function of its mass as well as the constraints provided by CMB and EGB observations and for different values of the duration (number of $e$-foldings) of reheating. As the figure shows, for some values of $N_{\rm reh}$ the cosmological restrictions could be violated for the range of masses of PBHs obtained from Model II. In particular, the green solid lines coincide with the minimum values that reheating had to last for Model II not to violate the CMB and EGB restrictions. The numerical values that we found were $N_{\rm reh} = 52$ and $N_{\rm reh} = 44$ for CMB anisotropies and EGB, respectively. Again, these values should be taken as lower bounds on how long reheating should last in order to agree with observational constraints from CMB anisotropies or EGB. Of course, this result is only valid for the parameters that we use for this work, so allowing other values of the model parameters is a task left for future work.
\section{Conclusions}\label{secVII}

In this article we have tested the inflationary potentials given in Eqs. \eqref{eq:pot1} (Model I) and \eqref{eq:pot2} (Model II) against a new mechanism of PBH formation proposed in P1. Once we calculate the primordial power spectrumn, we identify in both models a peak at small scales similar to the peak we proposed in P2. In particular, the peak in Model I occurs at scales closer to the end of inflation than in Model II. This brings consequences when evaluating for PBH formation through our new mechanism. For example, we obtained that for  Model I we should expect the formation of PBHs due to the gravitational collapse of Inflaton Haloes and Inflaton Stars if reheating lasts for at least $N_{\rm reh}= 21$ $e$-folds. Since the threshold value for the formation of PBHs due to the collapse of ISs is smaller than in the case of IHs, our results show a significant population of PBHs in model I only due to the ISs collapse. The mass in which these PBHs should have formed is of around $\sim 10^{4}-10^{5}~\rm{g}$. On the other hand, in Model II we have that the peak in the PPS occurs at scales not so close to the end of inflation, which means that ISs at the scales associated with the peak of the PPS cannot be formed. This has the consequence that we would only expect in this model the formation of PBHs due to the collapse of IHs with a mass of around $\sim 10^{13}-10^{14}~\rm{g}$. 

We have confronted both Models with the restrictions on PBHs that exist for the range of masses that formed in each of our models and we have found that if reheating doesn't last long enough, there would be a violation of the bounds provided by observations. Of course, although the conclusions given in this paper apply only to the parameter values given in the Tables \ref{tab:1} and \ref{tab:2}, and bound the duration of reheating. However, the general conclusions given in this work should continue to be valid for different parameters and, in general, for different inflationary models. Thus, the present work illustrates how to apply our new PBH formation criteria to other types of inflationary models.

\begin{acknowledgments}
{The authors  acknowledge support from program UNAM-PAPIIT, grants IN107521 “Sector Oscuro y Agujeros Negros Primordiales” and IG102123 ``Laboratorio de Modelos y Datos (LAMOD) para proyectos de Investigación Científica: Censos Astrofísicos". LEP and JCH acknowledge sponsorship from CONACyT
Network Project No.~304001 ``Estudio
de campos escalares con aplicaciones en cosmolog\'ia y
astrof\'isica'', and through grant CB-2016-282569.  The work of
LEP is also supported by the DGAPA-UNAM postdoctoral grants program, by CONACyT M\'exico under grants  A1-S-8742, 376127 and
FORDECYT-PRONACES grant No. 490769.}
\end{acknowledgments}

\bibliographystyle{ieeetr}
\bibliography{biblio}
\end{document}